**Determining the Location and Cause of Unintentional Quantum Dots in a Nanowire**

Ted Thorbeck[1,2], Neil M. Zimmerman[2]

1. Joint Quantum Institute & Department of Physics, University of Maryland, College Park, MD

2. Atomic Physics Division, NIST, Gaithersburg, MD

tedt@nist.gov

**Abstract:** We determine the locations of unintentional quantum dots (U-QDs) in a silicon nanowire with a precision of a few nanometers by comparing the capacitances to multiple gates with a capacitance simulation. Because we observe U-QDs in the same location of the wire in multiple devices, their cause is likely to be an unintended consequence of the fabrication, not random atomic-scale defects as is typically assumed. The locations of the U-QDs appear consistent with conduction band modulation from strain from the oxide and the gates. This allows us to suggest methods to reduce the frequency of U-QDs.

I. INTRODUCTION & MOTIVATION

Confining electrons in quantum dots enables the electrical manipulation and measurement of a single electron spin, which is a fundamental tool for quantum information and spintronics[1]. For these applications long electron spin lifetimes and coherence times are essential. In gallium arsenide quantum dots these times are limited by hyperfine coupling to the nuclear spins and spin-orbit coupling to the phonon bath[1]. This has motivated many efforts to electrically define quantum dots in silicon[2–6] and carbon[7–12], which are expected to have longer electron spin lifetimes and coherence times for two reasons. 1) Hyperfine coupling is reduced because only a few percent of nuclei have a spin, and isotopic enrichment can further reduce it. 2) The spin orbit coupling is reduced because the atomic masses are smaller.

Early work observing quantum dots in silicon[13,14] carbon nanotubes[7], and graphene nanoribbons[8,9] relied on uncontrolled potential modulation to define the quantum dot. While the causes of such potential modulations are typically unidentified, it is often assumed to be atomic-scale defects such as dopants[15], interface traps[4], kinks[16] or surface roughness[3]. Even though local gates controllably create the potential modulation that defines the quantum dots in more recent experiments, uncontrolled potential modulation has not been eliminated so additional unintentional quantum dots (U-QDs) are often observed[3–5,11,12].

In this paper, we describe a method to determine the location of U-QDs, by comparing measured gate capacitances to a simulation. Small uncertainties in the gate capacitances allow us to determine the location of the U-QDs with a precision of a few nanometers. Since we observe U-QDs in the same locations of a silicon nanowire in multiple devices, their cause is likely to be a systematic, but unintended, consequence of the fabrication, as opposed to random atomic-scale defects. The location of our U-QDs appears to be consistent with conduction band modulation from strain due to the oxide and the gates. After identifying a potential mechanism causing a U-QD, we can propose methods to reduce the frequency of U-QDs in devices. Our future task will be to experimentally verify that the proposed method actually reduces the frequency of U-QDs. We begin by discussing the devices and both intentional and unintentional quantum dots in sections II and III. Then in section IV, we devise a circuit model for the U-QDs and extract the gate capacitances. In section V and VI we determine the locations and causes of the U-QDs. We conclude in section VII.

II. DEVICE DETAILS

Our devices (Fig 1a and 1b) consist of a silicon nanowire surrounded by two poly-silicon gate layers. The wire is mesa etched on a lightly boron doped ($\approx 6 \times 10^{14}$ cm$^{-3}$) SOI wafer. The wire is protected by 20 nm of thermally grown SiO$_2$. Two layers of heavily doped poly-silicon gates wrap around the wire. The lower gate (LG) layer consists of three independent local gates: LGS, LGC and LGD (lower gate source, center and drain). The upper gate (UG) layer is a global gate that covers the entire device. Data from two devices will be presented in this paper. Device 1 (2) has a wire 530 (470) nm long, 17 nm thick, 20 (10) nm wide, and the LGs are 40 (40) nm long with 130 (100) nm in between LGs. A positive voltage on the UG inverts the wire. Negative voltages on the LGs locally deplete the wire directly below the gate; when the voltage is sufficiently negative a tunnel barrier is formed[2].

### III. INTENTIONAL AND UNINTENTIONAL QUANTUM DOTS

To create an intentional quantum dot, two of the LGs must be used to form tunnel barriers. When operating a single LG, making the voltage more negative will create a larger tunnel barrier, which typically turns current off smoothly (see the LGS trace in fig. 1c). For some of our LGs, the current undergoes a series of peaks before shutting off (see the LGD trace in fig 1c). The peaks are due to Coulomb blockade through U-QDs. These U-QDs are a common feature in our devices; in a sample of 8 devices we have observed similar structure in I($V_{LG}$) in 11 out of 24 LGs.

To get a better look at the U-QDs, we measure the current when scanning both the UG and LG, as shown in Fig. 2a for device 1 and in Fig. 2b for device 2. Notice that in both data sets only one LG is creating a tunnel barrier, so we have no intentional QDs. The striking similarity of these two datasets suggests that the U-QD have a non-random cause. In both figures two sets of parallel peaks appear; each set corresponds to Coulomb blockade peaks through a U-QD. As labeled in Fig. 2a, dot A causes the more steeply sloping set of peaks and dot B causes the less steeply sloped set of peaks. Where the two sets of peaks intersect, we observe an increase in current that is much larger than the sum of the current through the two peaks away from the crossing, as can be seen in Fig 2c.

Several features stand out about these two datasets. The two dots have very different characteristics: dot A experiences only a few peaks, whereas dot B experiences many peaks. Furthermore, the two dots appear to be interacting electrically; crossing a peak caused by dot A changes the current through dot B. For example, to the left of the leftmost peak of dot A (with the red outline in Fig 2a) there are no peaks corresponding to dot B.

### IV. CIRCUIT MODEL AND EXTRACTED PARAMETERS

We will use the capacitances from the gates to the U-QDs to determine the locations of the U-QDs. The capacitances can be deduced from the spacings and slopes of the peaks once we have a circuit model for the two quantum dots. We observe current through dots A and B individually, and a peak in current where the two dots intersect. Neither a single quantum dot nor a parallel double quantum dot (in which source and drain are tunnel coupled to only one dot) can produce two sets of peaks with different slopes. On the other hand, a series double quantum dot can explain the current where the two dots intersect but cannot explain the current through each dot individually. Finally, cotunneling in the series double quantum dot can qualitatively explain both the peaks where the two dots intersect as well as each dot individually. However, when we attempted to simulate our measurement with cotunneling in a series double quantum dot, no set of resistances could explain the magnitude and ratios of current in the peaks, so this model was rejected.

The circuit model (Fig. 3a) that best explains the data has a series path through both dots (Source-1-A-2-B-3-Drain) and two parallel paths each through one dot (Source-1-A-5-Drain; Source-4-B-3-Drain). The more steeply sloped set of peaks corresponds to the parallel path through dot A. The less steeply sloped set of peaks corresponds to the parallel path through dot B. A lower resistance series path through both A and B is needed to explain why there is more current when the two lines intersect than through either line individually (Fig 2C). This circuit model has not been reported in previous experiments. In the next few paragraphs, we will extract the parameters for the circuit model by measuring the data and comparing the data to a simulation of our circuit model.

The slope and spacing ($\Delta V_G$) of peaks in Figs. 2a and 2b allow us to measure the capacitance ($C_G = e\ \Delta V_G$) to both the upper gate ($C_{UG}$) and lower gate ($C_{LG}$). Comparing the data in Figs 2a and 2b to simulations as in Fig 3b allows us to deduce the resistances and capacitances of the tunnel junctions. The circuit model was simulated by using a single-electron circuit simulator[17,18]. The resistance of the parallel path through dot B depends on the charge state of dot A (the current in the less steeply sloped set of peaks changes when crossing a peak of the more steeply sloped set). We accounted for this by multiplying the simulation results by an envelope function taken at high $V_{UG}$ where there was no Coulomb blockade through dot B but changes in the charge state of dot A could be clearly seen. The results of the simulation are shown in Fig. 3b for the parameters of device 1 in Tables I and II. The simulation matches the data nicely, giving us confidence that the circuit model and its parameters are correct.

Table 1 contains the gate capacitances used in the circuit model and table 2 contains the resistances and capacitances of the tunnel junctions[19]. The relative uncertainty of the ratio of capacitances ($C_{LGD(S)}/C_{UG}$) in Table 1 is smaller than the relative uncertainty of the capacitances. In dot A in device 2, for example, the relative uncertainty of the capacitance was large because we only observe a few peaks. In contrast, the uncertainty of the ratio is small because the slope of the peak was consistent over 2 V of $V_{UG}$, allowing a very precise measurement.

V. DETERMINING THE LOCATION OF THE U-QDS

We will determine the size and location of the U-QDs by comparing the gate capacitances to a simulation. We use the gate capacitances rather than the tunnel junction resistance and capacitance to locate the dot because the gate capacitance is insensitive to gate voltage[2]. First, we can use some intuition to roughly locate the dots. In both devices dot A is more strongly coupled to the LG than to the UG, so dot A is probably closer to the LG. However, there is still a substantial coupling to the UG so dot A is probably not located directly underneath the LG. Therefore, the best guess for the location of dot A is in the gap between the UG and the LG, closer to the LG. Dot B is more strongly coupled to the UG than to the LG; however, dot B has larger capacitances to both gates than dot A. This suggests that dot B is a much larger dot; although part of it is close to the LG it likely extends a long way directly under the UG.

We used FASTCAP (an electromagnetic field solver)[18,20] to simulate the gate capacitances, as validated by comparison to intentional dots measurements[21]. By taking the known dimensions of the wire and gates and then slicing the wire up into narrow slices we can simulate the differential gate capacitances ($dC/dz$), where z is the position along the wire. The slices are assumed to wrap around all four sides of the wire. For additional discussion of capacitance simulation in these devices see ref 21. Integrating the differential capacitance $\int_{z_1}^{z_2} \frac{dC}{dz} dz$ over all possible start ($z_1$) and end ($z_2$) points of the dot and keeping only those that are consistent with our measured gate capacitances, we can determine the size and location of the dots. The simulated differential capacitance for the drain half of device 1 is shown in Fig.

4a; as expected $dC_{LGC}/dz$ and $dC_{LGD}/dz$ are strongly peaked directly below their respective gates and $dC_{UG}/dz$ is larger away from the LGs.

The range of sizes and locations in the simulation that reproduce our measured capacitances are in table 3 and shown in Figs. 4a. and Fig. 4b. Because our device look roughly symmetric about LGD (LGS for device 2), there were initially two locations that satisfied all three capacitance conditions from table I, one on either side of LGD. To resolve this symmetry, a separate measurement showed that the capacitive coupling to LGC was very small (< 0.1 aF). Of the two possible locations for dot B, only one satisfied this new condition on the capacitance to LGC (the location further from LGC). For dot B in device 2, we were initially unable to find any location or size for a dot that was consistent with our measurement. We solved this by assuming the LGs were fabricated 30 nm off-center, which is sometimes seen in fabrication. Notice the similarity in the location of A and B in the two devices, and that both dots are consistent with our intuitive guesses for the locations. The fact that we can self-consistently predict the values of both $C_{UG}$ and $C_{LG}$ gives us confidence in this technique. Additional simulations (not shown) show that small variations in the length of the gate or thickness of the oxide do not change the deduced location or size of the dots, giving us additional confidence in our results.

The uncertainties on $z_{1,2}$ give the range over which the integrated capacitance is within the experimental uncertainty in the gate capacitance. Estimating the error propagation from the experimental gate capacitance (ΔC) to the position ($\Delta z_{1,2}$) as $\Delta z_{1,2} \approx \frac{\Delta C}{dC\ dz_{1,2}}$, shows that a precision on the location of a few nanometers is reasonable. In dot B of device 1, for example, the uncertainty in the position ($z_2$ = 87

± 2 nm) can be estimated as $\Delta z_2 \approx 3$ nm using $\Delta C_{UG}$ = 0.3 aF (from experiment) and $dC/dz_2$ = 0.12 aF/nm (from simulation).

## VI. DISCUSSION OF THE CAUSES OF THE U-QDS

Knowing the location of the two U-QDs, we can now discuss possible causes of the U-QDs. Because both dots A and B are in the same location in both devices, we suspect the cause of the dots is some unintentional but systematic consequence of the fabrication. We suggest strain, from either the oxide or the gates, is a plausible cause[22]. This suggestion is in contrast to the typical assumption that U-QDs are caused by random atomic defects such as interface traps. Both the oxide and the gates can cause strain in the wire, which will alter the bandgap of silicon[23,24]; a local dip in the conduction band could cause a quantum dot.

The likely cause of dot B is i) an intentional tunnel junction caused by negative voltages on $V_{LG}$ and ii) an unintentional strain-induced tunnel junction at the end of the nanowire. Silicon nanowires without LGs[25] have shown Coulomb blockade behavior because of naturally occurring tunnel junctions at the ends of the nanowire. These tunnel junctions are caused by a decrease in oxide-induced strain at the ends of the wire, because they are less exposed to the thermal oxidation[23]. Understanding the cause of dot B, we can now suggest methods to eliminate the dots from future devices. 1) Increasing the distance from the lower gates to the ends of the wire would lower the charging energy of the U-QD below the temperature, where the dot will no longer show Coulomb blockade. 2) Adding a separate UG for the source and drain portions of the wire would position the Fermi level above the peak in the conduction band causing the tunnel junction. Such simple steps could drastically reduce the frequency of U-QDs in our devices.

Although we know the location of dot A, we need to do additional work to understand its origin. Previous work on strain in finFETs (similar to our devices but with no UG and only one LG) has shown that the gate can lead to compressive strain in the channel[24]. Since strain modulation can lead to modulation of the conduction band[23], we suspect that this is the origin of dot A. More work is needed to verify this for our geometry and to suggest methods to eliminate this mechanism.

## VII. CONCLUSIONS AND ACKNOWLEDGEMENTS

We have shown how the location and size of unintentional quantum dots can be determined to within several nanometers by a multiple gate capacitance measurement. The capacitances were extracted from the measurements and verified by a comparison to a circuit simulation. The similar nature of the U-QDs in both devices suggests a common origin, which is unlikely to be a random atomic-scale defect as is typically assumed in the field. Instead, strain-induced modulation of the bandgap was suggested as the origin of the dots. Two possible ways of eliminating one of the dots was suggested. The commonly observed U-QDs in other geometries and materials, as discussed in the introduction, might also be unintentional consequences of the fabrication, whether it is through strain-induced conduction band modulation or some other method. Since gate capacitances are routinely measured this technique is a simple but powerful method to help determine the cause of U-QDs and to reduce their frequency.

We gratefully acknowledge the fabrication efforts of Akira Fujiwara, Yukinori Ono, Hiroshi Inokawa, and Yasuo Takahashi. We would also like to acknowledge helpful conversations with Michael Stewart, Josh Pomeroy, Panu Koppinen, Russell Lake and Curt Richter. This research was supported by the Laboratory for Physical Sciences (EAO93195).

FIG. 1

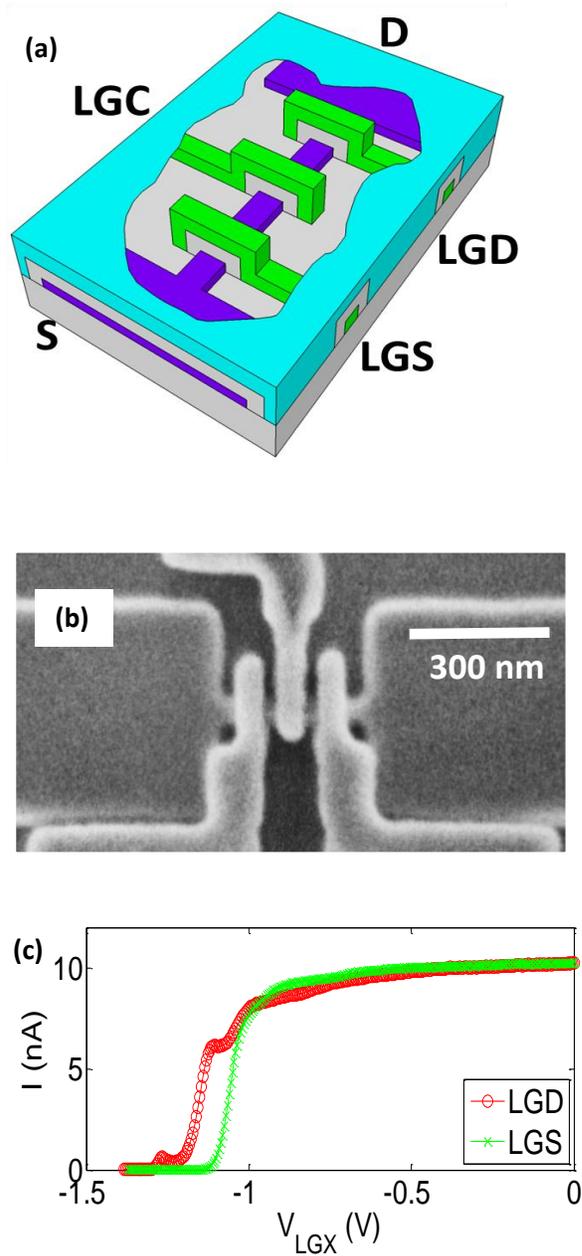

FIG. 1. (a) Cut-away cross section of the device. The silicon nanowire is in purple, silicon dioxide is in grey, the LGs are in bright green and the UG (not labeled) is turquoise. (b) SEM of the device before encapsulation by UG. (c) Comparison of $V_{LGS}$ and $V_{LGD}$ scan taken in device 1 at $V_{UG}$ = 2 V, T = 4.2 K and $V_{SD}$ = 1 mV. Oscillations due to U-QD are seen for LGD.

FIG. 2

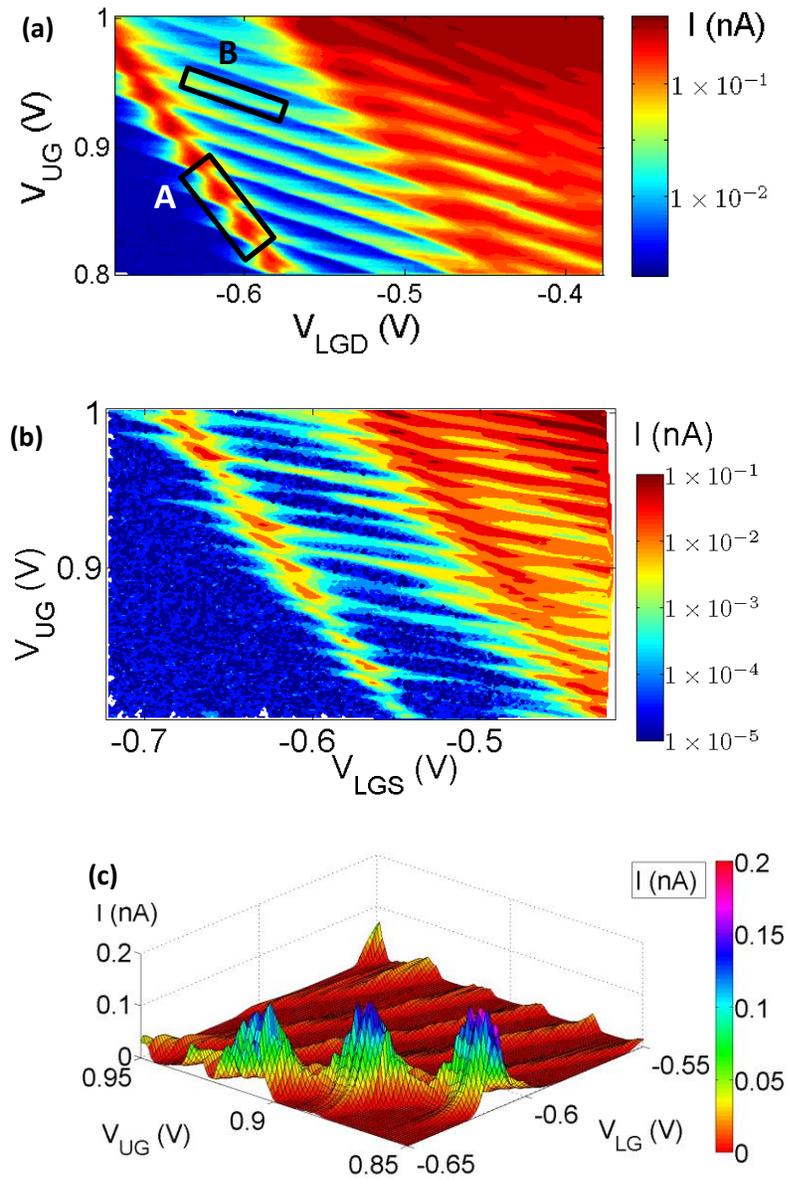

FIG. 2. (a) and (b) Current measurement as both UG and LG are scanned for device 1 (a) and device 2 (b) for $V_{SD}$ = 1 mV. Box "B" contains a peak due to current through dot B, and box "A" contains both a more steeply-sloped peak due to current through dot A, as well as two intersections with the peaks due to current through dot B. Both were measured at the base temperature of a dilution

refrigerator. (c) Psuedo-3D view of data from (a), highlighting the much larger current where A and B cross.

FIG. 3

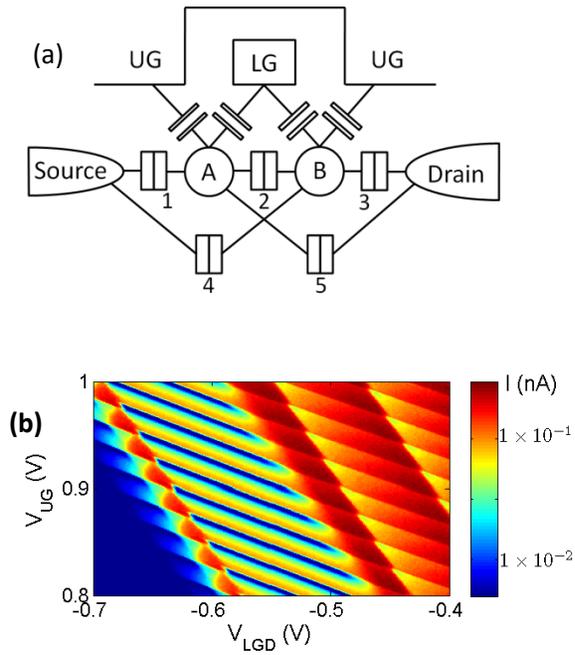

(a)

(b)

FIG. 3. (a) Circuit diagram used in this paper. The tunnel junctions are labeled 1 to 5. The horizontal path represents the series path (S-1-A-2-B-3-D) which has the least resistance. The two tunnel junctions below the dots represent the higher resistance parallel paths (S-4-B-3-D and S-1-A-5-D). (b) The results of a simulation of the current through the circuit model (a) using parameters in tables 1 and 2 for device 1.

FIG. 4

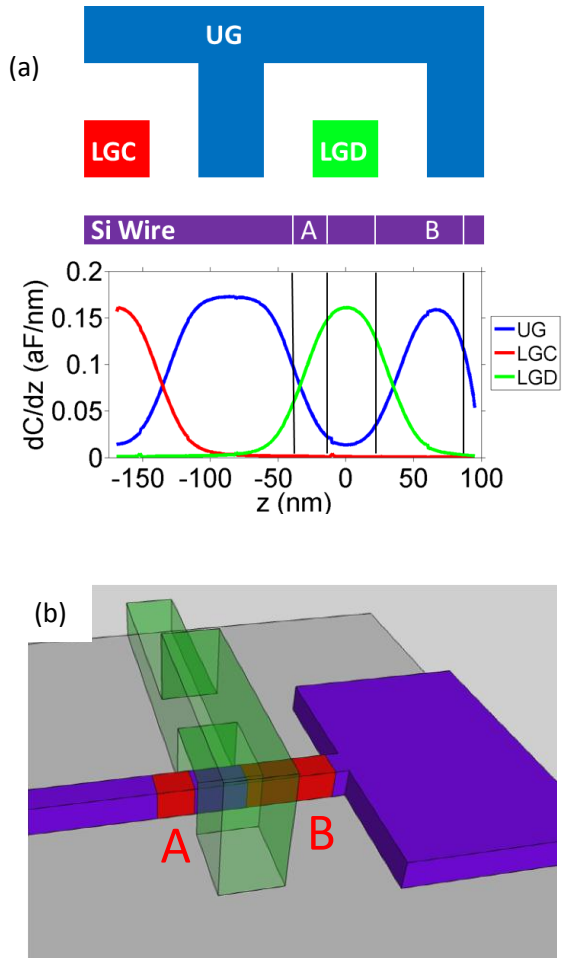

Fig. 4. (a) The top half the shows a cross section of device 1 with the Si wire, LGC, LGD and UG shown to scale along the drain half of the device. The bottom half shows the differential capacitance for device 1 from the gates to the wire (z-axis origin is the center of LGD). The vertical lines in the Si wire and in the capacitance graph show the calculated range of the two U-QDs. (b) A pseudo-3d view of the wire with the U-QD in red and the LG translucent.

Table 1

| Dev. 1 | LG (aF) | UG (aF) | Slope | Dev. 2 | LG (aF) | UG (aF) | Slope |
|---|---|---|---|---|---|---|---|
| A | 2.3 + 0.3 - 1.3 | 1.3 + 0.2 - 0.6 | -1.71 ± 0.02 | A | 1.3 ± 0.2 | 0.9 ± 0.1 | -1.46 ± 0.03 |
| B | 3.2 ± 0.2 | 7.9 ± 0.3 | -0.41 ± 0.01 | B | 2.2 ± 0.2 | 12.2 ± 0.6 | -0.179 ± 0.007 |

Table 1 Gate capacitance from LG and UG to dots A and B as well as the slope of the lines for both devices, as measured from the data in Fig 2. Uncertainties in capacitances and slopes represent the maximum and minimum spacing and slopes that can be fit given the widths of the peaks.

Table 2

| Dev. 1 | R (Ω) | C (aF) | Dev. 2 | R (Ω) | C (aF) |
|---|---|---|---|---|---|
| 1 | 100 k | 15 | 1 | 1 M | 20 |
| 2 | 3 M | 10 | 2 | 15 M | 10 |
| 3 | 100 k | 10 | 3 | 1 M | 35 |
| 4 | 10 M | 100 | 4 | 500 M | 60 |
| 5 | 6 M | 10 | 5 | 65 M | 10 |

Table 2 Complete tunnel junction parameters for both devices as labeled in figure 3. Parameters were deduced from comparing simulations, as in Fig 2c, to measurements, as in Figs 2a and 2b. Since the full range of parameters that adequately reproduce the measured current was not explored, uncertainties for this table were not evaluated.

Table 3

| Dev. 1 | $z_1$ (nm) | $z_2$ (nm) | Dev. 2 | $z_1$ (nm) | $z_2$ (nm) |
|---|---|---|---|---|---|
| A | -40 ± 3 | -19 ± 3 | A | -37 ± 1 | -22 ± 1 |
| B | 17 ± 1 | 87 ± 2 | B | 23 ± 2 | 117 ± 3 |

Table 3 Bounds of both dots in both devices. Position is defined using z-axis defined in Fig 4a, where z = 0 is the center of the LG and the ends of the LG are at ± 20 nm and the UG starts at ± 50 nm. Notice the similarities in the locations for dot A and B in the two devices.

References

[1] R. Hanson, L.P. Kouwenhoven, J.R. Petta, S. Tarucha, and L.M.K. Vandersypen, Rev. Mod. Phys. **79**, 1217 (2007).

[2] A. Fujiwara, H. Inokawa, K. Yamazaki, H. Namatsu, Y. Takahashi, N.M. Zimmerman, and S.B. Martin, Applied Physics Letters **88**, 053121 (2006).

[3] H.W. Liu, T. Fujisawa, Y. Ono, H. Inokawa, A. Fujiwara, K. Takashina, and Y. Hirayama, Phys. Rev. B **77**, 073310 (2008).

[4] E.P. Nordberg, G.A.T. Eyck, H.L. Stalford, R.P. Muller, R.W. Young, K. Eng, L.A. Tracy, K.D. Childs, J.R. Wendt, R.K. Grubbs, J. Stevens, M.P. Lilly, M.A. Eriksson, and M.S. Carroll, Phys. Rev. B **80**, 115331 (2009).

[5] B. Hu and C.H. Yang, Phys. Rev. B **80**, 075310 (2009).

[6] S.J. Angus, A.J. Ferguson, A.S. Dzurak, and R.G. Clark, Nano Lett. **7**, 2051 (2007).

[7] M. Bockrath, D.H. Cobden, P.L. McEuen, N.G. Chopra, A. Zettl, A. Thess, and R.E. Smalley, Science **275**, 1922 (1997).

[8] K. Todd, H.-T. Chou, S. Amasha, and D. Goldhaber-Gordon, Nano Lett. **9**, 416 (2008).

[9] C. Stampfer, J. Güttinger, S. Hellmüller, F. Molitor, K. Ensslin, and T. Ihn, Phys. Rev. Lett. **102**, 056403 (2009).

[10] M.J. Biercuk, S. Garaj, N. Mason, J.M. Chow, and C.M. Marcus, Nano Lett. **5**, 1267 (2005).

[11] X.L. Liu, D. Hug, and L.M.K. Vandersypen, Nano Lett. **10**, 1623 (2010).

[12] N. Mason, M.J. Biercuk, and C.M. Marcus, Science **303**, 655 (2004).

[13] T.E. Kopley, P.L. McEuen, and R.G. Wheeler, Phys. Rev. Lett. **61**, 1654 (1988).

[14] J.H.F. Scott-Thomas, S.B. Field, M.A. Kastner, H.I. Smith, and D.A. Antoniadis, Phys. Rev. Lett. **62**, 583 (1989).

[15] Y. Ono, K. Nishiguchi, A. Fujiwara, H. Yamaguchi, H. Inokawa, and Y. Takahashi, Applied Physics Letters **90**, 102106 (2007).

[16] M.J. Biercuk, N. Mason, J.M. Chow, and C.M. Marcus, Nano Lett. **4**, 2499 (2004).

[17] C. Wasshuber, *Computational Single-Electronics*, 1st ed. (Springer, 2001).

[18] (Software is named for informational purposes only; it does not imply an endorsement or a recommendation by NIST).

[19] Since the more steeply sloped lines in Fig 2a are not periodic, we assume the smaller spacing gives us the capacitance, and we attribute the larger spacing to an additional few electron charging energy. This leads us to assign a large and asymmetric uncertainty to the corresponding gate capacitances in Table 1.

[20] K. Nabors and J. White, IEEE Transactions on Computer-Aided Design of Integrated Circuits and Systems **10**, 1447 (1991).

[21] T. Thorbeck and N. Zimmerman, Forthcoming.

[22] T. Thorbeck and N. Zimmerman, Forthcoming.

[23] Y. Ono, K. Yamazaki, M. Nagase, S. Horiguchi, K. Shiraishi, and Y. Takahashi, Solid-State Electronics **46**, 1723 (2002).

[24] Kyoungsub Shin, Chi On Chui, and Tsu-Jae King, in *Electron Devices Meeting, 2005. IEDM Technical Digest. IEEE International* (IEEE, 2005), pp. 988-991.

[25] Y. Takahashi, Y. Ono, A. Fujiwara, and H. Inokawa, Journal of Physics: Condensed Matter **14**, R995 (2002).